\documentclass[10pt, conference]{IEEEtran}

\newif\ifSPACEHACK
\SPACEHACKtrue 

\newif\ifDEBUG
\DEBUGtrue

\newif\ifANONYMOUS
\ANONYMOUSfalse

\usepackage{algorithmic}
\usepackage{graphicx}
\usepackage{textcomp}
\usepackage{xcolor}
\usepackage{booktabs} 
\usepackage{xspace} 
\usepackage[normalem]{ulem}
\usepackage{makecell}
\usepackage{tcolorbox}
\usepackage{enumitem}
\usepackage{siunitx}
\usepackage[vskip=1em,font=itshape,leftmargin=2em,rightmargin=2em]{quoting}
\usepackage[compact]{titlesec}
\usepackage{tabularx}
\usepackage{hyperref}
\usepackage{lscape}
\usepackage{afterpage}

\usepackage{makecell}
\usepackage{multirow}
\usepackage{booktabs}
\usepackage[T1]{fontenc}
\usepackage{stfloats}

\usepackage{filecontents,lipsum}
\usepackage[noadjust]{cite}

\usepackage{soul}

\ifDEBUG
    \newcommand{\JD}[1]{\textcolor{purple}{[JD:#1]}}
    \newcommand{\AG}[1]{\textcolor{olive}{[AG:#1]}}
    \newcommand{\WJ}[1]{\textcolor{teal}{[WJ:#1]}}
    \newcommand{\GKT}[1]{\textcolor{brown}{[GKT:#1]}}
    \newcommand{\NMS}[1]{\textcolor{orange}{[NMS: #1]}}
    \newcommand{\NV}[1]{\textcolor{red}{[NV: #1]}}
    \newcommand{\JY}[1]{\textcolor{cyan}{[JY: #1]}}
    \newcommand{\JJ}[1]{\textcolor{violet}{[JJ: #1]}}
    \newcommand{\YT}[1]{\textcolor{blue}{[YT: #1]}}

    \newcommand{\TODO}[1]{\hl{#1}}
\else
    \newcommand{\JD}[1]{}
    \newcommand{\AG}[1]{}
    \newcommand{\WJ}[1]{}
    \newcommand{\GKT}[1]{}
    \newcommand{\NMS}[1]{}
    \newcommand{\NV}[1]{}
    \newcommand{\JY}[1]{}
    \newcommand{\JJ}[1]{}
    \newcommand{\YT}[1]{}

    \newcommand{\TODO}[1]{\hl{#1}}
\fi

\usepackage[compact]{titlesec}
    \usepackage{etoolbox}
    \makeatletter
    \patchcmd{\ttlh@hang}{\parindent\z@}{\parindent\z@\leavevmode}{}{}
    \patchcmd{\ttlh@hang}{\noindent}{}{}{}
    \makeatother

\newcommand{\myparagraph}[1]{\vspace{0.07cm}\noindent\textbf{#1} \noindent{}}

\ifSPACEHACK
    \titlespacing*\section{0pt}{4pt plus 1pt minus 1pt}{3pt plus 1pt minus 1pt}
    \titlespacing*\subsection{0pt}{4pt plus 1.5pt minus 1.5pt}{3pt plus 1.5pt minus 1.5pt}
    \titlespacing*\subsubsection{0pt}{3pt plus 1pt minus 1pt}{3pt plus 1.5pt minus 1.5pt}
    \titlespacing*\paragraph{0pt}{2pt plus 1.5pt minus 1.5pt}{2pt plus 1.5pt minus 1.5pt}
\fi

\usepackage{balance} 
\usepackage[font=small,labelfont=bf]{caption}
\usepackage{subcaption}
\makeatletter
\def\cl@chapter{}
\makeatother
\usepackage{cleveref}

\crefformat{section}{\S#2#1#3}
\crefname{figure}{Figure}{Figures}
\crefname{appendix}{Appendix}{Appendices}
\crefname{table}{Table}{Tables}
\crefname{algorithm}{Algorithm}{Algorithms}
\crefname{listing}{Listing}{Listings}
\crefname{theorem}{Theorem}{Theorems}
\crefname{thm}{Theorem}{Theorems}
\crefname{lemma}{Lemma}{Lemmata}
\crefname{equation}{Eqt.}{Eqts.}
\crefformat{Grammar}{Grammar #1}

\usepackage{enumitem}
\usepackage{booktabs}

\newcommand{\ie}{\textit{i.e.,} }
\newcommand{\eg}{\textit{e.g.,} }
\newcommand{\etal}{\textit{et al.}\xspace}

\newcommand{\code}[1]{{\small\texttt{#1}}\xspace}

\newcommand{\DatasetNickname}{\emph{PeaTMOSS}\xspace}
\newcommand{\DatasetNicknameNormal}{PeaTMOSS\xspace}
\newcommand{\DatasetNicknameFormatted}{\code{\textbf{P}ea\textbf{TMOSS}}\xspace}
\newcommand{\DatasetNicknameExpanded}{\textbf{\ul{P}}re-\textbf{\ul{T}}rained \textbf{\ul{M}}odels in \textbf{\ul{O}}pen-\textbf{\ul{S}}ource \textbf{\ul{S}}oftware\xspace}

\newcommand{\NEWnumberOfModelHub}{{2}\xspace}
\newcommand{\NEWnumberOfPTMs}{{281,638}\xspace}

\newcommand{\NEWnumberOfPTMRepos}{{15,250}\xspace}
\newcommand{\NEWDatasetSize}{{48.2 TB}\xspace}
\newcommand{\NEWSizeOfMetadata}{{7.12 GB}\xspace}
\newcommand{\NEWSizeOfRepos}{\TODO{xxx TB}\xspace}

\newcommand{\DBAccessExamplesPath}{{\code{/Examples/}}\xspace}

\newcommand{\DBFileName}{\code{PeaTMOSS.db}\xspace}

\newcommand{\TotalNumberOfLinks}{{44,337}\xspace}

\newcommand{\GitHubReuseSignaturePT}{{163}\xspace}

\newcommand{\GitHubReuseLibrariesHF}{{23}\xspace}

\newcommand{\GitHubReuseRepoCountSourceGraph}{{36,251}\xspace}
\newcommand{\GitHubReuseRepoCountTP}{{28,575}\xspace}
\newcommand{\GitHubReuseRepoCountStaticMatch}{{15,129}\xspace}
\newcommand{\GitHubReuseRepoCountInDB}{{27,270}\xspace}
\newcommand{\GitHubReuseRepoCountSourceGraphSize}{{3.5 TB}\xspace}

\newcommand{\GitHubReusePrCount}{{12,159}\xspace}
\newcommand{\GitHubReuseIssueCount}{{19,507}\xspace}

\newcommand{\GitHubReuseRepoSourceGraphAvgStar}{{201}\xspace}
\newcommand{\GitHubReuseRepoSourceGraphDate}{{July 10, 2023}\xspace}

\newcommand{\numberOfModelHub}{5\xspace}

\newcommand{\HFNumberOfPackagesRepos}{{15,913}\xspace}

\newcommand{\TotalNumberOfPackagesMetadata}{{\NEWnumberOfPTMs}\xspace}

\begin{document}


\renewcommand{\thefootnote}{\fnsymbol{footnote}} 

\newcommand{\DatasetGitHubURL}{\url{https://github.com/PurdueDualityLab/PeaTMOSS-Demos}}


\newcommand{\mytitle}{}
\renewcommand{\mytitle}{MS-DOS: Mining Software--Deep Learning in Open-Source}
\renewcommand{\mytitle}{MS-DOS: The Pre-Trained Models Mining Challenge of Deep Learning in Open-Source}
\renewcommand{\mytitle}{MS-DOS: The Mining Software Repositories Challenge for Pre-Trained Models---Deep Learning in Open-Source}
\renewcommand{\mytitle}{MS-(PTM)DOS: The Mining Software Repositories Challenge for Pre-Trained Models---Deep Learning in Open-Source}
\renewcommand{\mytitle}{PeatMOSS: The Mining Challenge for PrE-Trained Models in Open-Source Software}
\renewcommand{\mytitle}{PTMChain: Mining Pre-Trained Models in the Open-Source Software Supply Chain}
\renewcommand{\mytitle}{\DatasetNicknameNormal: Mining Pre-Trained Models in Open-Source Software}

\title{\mytitle}

\ifANONYMOUS
  \author{Anonymous Author(s)}
\else
    \vspace{-1cm}
    \author{
    \IEEEauthorblockN{Wenxin Jiang\footnotemark$^{1 ^ *}$,
    Jason Jones\footnotemark$^{1 ^ *}$,
    Jerin Yasmin\footnotemark$^{2 ^ *}$,
    Nicholas Synovic$^{3}$,
    Rajeev Sashti$^{1}$,
    Sophie Chen$^{4}$,\\
    George K. Thiruvathukal$^{3}$,
    Yuan Tian$^{2}$,
    James C. Davis\footnotemark$^{1}$}
    \IEEEauthorblockA{Purdue University$^{1}$ ; Queen's University$^{2}$ ; Loyola University--Chicago$^{3}$ ; and University of Michigan--Ann Arbor$^{4}$}
      {West Lafayette, Indiana, USA$^{1}$ ; Kingston, Ontario, CA$^{2}$ ; Chicago, Illinois, USA$^{3}$ ; and Ann Arbor, Michigan, USA$^{4}$}
    }
\fi



\maketitle

\begin{abstract}
Developing and training deep learning models is expensive, so
software engineers have begun to reuse pre-trained deep learning models (PTMs) and fine-tune them for downstream tasks.
Despite the wide-spread use of PTMs, we know little about the corresponding software engineering behaviors and challenges.

To enable the study of software engineering with PTMs, we present the \DatasetNicknameFormatted dataset: \DatasetNicknameExpanded.
\DatasetNickname has three parts:
  a snapshot of
    (1) \NEWnumberOfPTMs PTMs,
    (2) \GitHubReuseRepoCountInDB open-source software repositories that use PTMs,
    and
    (3) a mapping between PTMs and the projects that use them. 
We challenge \DatasetNickname miners to discover software engineering practices around PTMs.
A demo and link to the full dataset are available at: \DatasetGitHubURL.

\end{abstract}

\footnotetext[1]{These authors contributed equally and are listed alphabetically.}

\vspace{-0.1cm}
\section{High-Level Overview}
\vspace{-0.1cm}


\myparagraph{Motivation:}
Deep Neural Networks (DNNs) have become a common component in software systems over the past decade. 
While some software engineers develop DNNs from scratch, many others integrate DNNs into their software following a typical re-use pattern:
  (1) pre-trained DNN models are published to registries such as Hugging Face, similar to traditional software package registries (\eg NPM, PyPI);
  and 
  (2) other software depends on these Pre-Trained Models (PTMs), accessed via libraries or web APIs.

\vspace{0.08cm}
\emph{\textbf{Despite wide-spread adoption of PTMs, we know relatively little about how PTMs are integrated into software systems.}}
\vspace{0.05cm}



\begin{figure}[h!]
    \centering
    \includegraphics[width=0.90\linewidth]{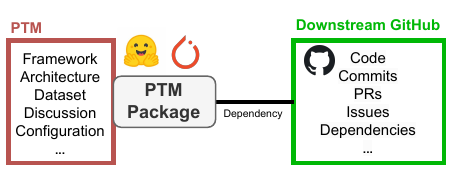}
    \caption{Data for \DatasetNicknameExpanded.}
    \label{fig:Overview}
\end{figure}

\myparagraph{Challenge:}
\hspace{-1em}
We propose the \DatasetNickname challenge to learn more about \DatasetNicknameExpanded (\cref{fig:Overview}).

The \DatasetNickname dataset comprises snapshots of PTMs and open-source repositories utilizing PTMs, as well as a mapping of PTMs to projects. 
\DatasetNickname contains
  \TotalNumberOfPackagesMetadata PTM packages,
  \GitHubReuseRepoCountInDB GitHub projects that use PTMs as dependencies,
  and \TotalNumberOfLinks links from these GitHub repositories to the PTMs they depend on. 
For both PTMs and GitHub projects, \DatasetNickname contains metadata
(commits, issues, pull requests) and data (\eg model architecture and weights; \code{git} repositories).
A uniform schema for retrieving PTM and project metadata is provided to assist in mining efforts.
Most information is indexed; some is stored as blobs.
The dataset can be accessed in two formats. 
The ``metadata'' version of \DatasetNickname is \NEWSizeOfMetadata and contains only the metadata of the PTM packages and a subset of the GitHub project metadata.
The ``full'' version is \NEWDatasetSize 
, adding 
    (1)
      the PTM package contents in each published version,
    and 
    (2)
      \code{git} history of the main branches of the GitHub projects.

\section{\DatasetNicknameNormal Dataset Structure}
\vspace{-0.05cm}


The metadata version of \DatasetNickname is stored in a SQLite database.
The tables include hyperlinks to tarred copies of the PTM package or GitHub repository. 
Dataset schemas are described in \cref{sec:Schema}.

The mapping between GitHub projects and PTMs are cases where a GitHub repository is known to depend on a particular PTM.
Additional detail is given in~\cref{sec:DataCollection-Linking}.

\section{Accessing and Working with \DatasetNicknameNormal}
\balance

\myparagraph{Access:}
The metadata and full versions of \DatasetNickname are available through a Globus share hosted at Purdue University.
\DatasetNickname can be downloaded via the official Globus Connect application, which is available for all major operating systems, \eg  Linux, MacOS, and Windows.
We include a Python script to download and configure an SQLite instance of the metadata version.
For more instructions, see \DatasetGitHubURL.

\myparagraph{Working with PeaTMOSS:}
To interact with \DatasetNickname, we recommend ORM or SQL.
Examples are provided in \cref{sec:Access_Examples}.

\myparagraph{Required Skills:}
\DatasetNickname is accessible to miners with a range of skills and interests.
\DatasetNickname includes both
  standard mining artifacts from GitHub (\eg \code{git} repositories, issues, PRs)
  and
  unusual artifacts from PTMs (\eg neural networks, weights, model cards).
Miners interested in program analysis, software repository mining, and natural language processing, etc. may apply these techniques to GitHub data, PTM data, or both.

\emph{Neither expertise in deep learning nor access to hardware such as GPUs is necessary for use of \DatasetNickname}.
Of course, miners with deep learning expertise or hardware resources could explore more advanced questions about PTM usage on GitHub or delve deeper into the PTM data.

\myparagraph{Data Samples:}
We offer a subset of samples from \DatasetNickname at: \DatasetGitHubURL.

\section{Possible Research Questions}
\balance

\cref{table:RQs} presents sample research questions for miners to investigate. 
This table includes questions focused on
  the GitHub portion of the dataset,
  on the PTM portion of the dataset,
  and
  on both parts.
It notes some prior work as starting points for miners.

\renewcommand{\arraystretch}{1.2}
{
\begin{table*}[hb!]
\centering
\caption{
Example questions for miners to investigate.
These questions are divided into three groups.
The first group of questions makes use of the GitHub portion of the dataset (\emph{GH}).
The second group uses the Pre-Trained Model portion of the dataset (\emph{PTM}).
The third group asks questions that require Integrating both parts of the dataset (\emph{I}).
}

\label{table:RQs}
\begin{tabular}{p{14cm} |p{2cm}}
\toprule
\textbf{Research questions} & \textbf{Related work} \\
\toprule


{\textbf{GH1:}} What kinds of defects are opened related to PTM use in the GitHub projects? How do these defects differ from defects opened on other aspects of the GitHub projects? & \cite{morovati2023buginMLSystems} \\

{\textbf{GH2:}} What do developers on GitHub discuss related to PTM use, \eg in issues, and  
pull requests? What are developers' sentiments regarding PTM use?
Do the people do pull requests of PTMs have the right expertise?
 & \cite{yin2020team, sajadi2023interpersonalTrustinOSS}
 \\
 
{\textbf{GH3:}} How commonly do developers change the specific PTM they use to implement a feature? What factors influence these changes? & \cite{dilhara2021understanding}
\\

{\textbf{GH4:}} Sometimes a PTM is introduced into a GitHub repository as part of the initial implementation of a software feature, but other times a PTM is introduced to replace part or all of an existing feature implementation. How common are these two modes of PTM adoption? In the second kind, how does the feature's defect types or defect rates change after the PTM is introduced? & \cite{li2021modeldiff, zhang2022ReMoS, qi2023reusingDNNthroughModelReengineering,nahar2023dataset} \\

\midrule

\textbf{PTM1:} What factors predict the popularity of a PTM, and what is their relative importance? Intuition suggests that performance aspects such as accuracy and latency may dominate; what is the role played by other factors such as engineering quality? & \cite{lima2020characteristics, Jiang2022PTMReuse}
\\

\textbf{PTM2:} Recent qualitative work determined that software engineers struggle to re-use PTMs because of their limited documentation. What are the typical characteristics of this documentation? Can natural-language model cards be automatically parsed into a structured schema? & \cite{Jiang2022PTMReuse, Jiang2022PTMSupplyChain,bhat2023aspirations} \\

\textbf{PTM3:} One aspect of re-use is finding a candidate model. What naming conventions do PTMs follow? Are they consistent enough (within an architecture family? across families?) to support engineers looking for similar models? When do PTM maintainers release a model under a new name, and when do they simply bump the version number? & \cite{gresta2023naminginOOP}
\\

\textbf{PTM4:} PTM authors may reuse each others' work, \eg building off of model checkpoints or incorporating architectural building blocks. This might be viewed as an extreme form of ``forking'' from open-source software, but it may also reflect a novel form of software exchange. What is the phylogeny, or perhaps the ``supply chain'', of the major families of PTMs? 
& \cite{Jiang2022PTMSupplyChain} \\

\textbf{PTM5:} Many research papers describe techniques for identifying DNNs with unexpected behavior, \eg hidden malicious behaviors. How common are such DNNs in the PTM dataset? 
& \cite{Jiang2022PTMReuse, Wang2022Backdoor4TLwithPTM, Wang2022EvilModel2, guo2022threats}\\

\midrule

\textbf{I1:} It can be difficult to interpret model popularity numbers by download rates. To what extent does a PTM's download rates correlate with the number of GitHub projects that rely on it, or the popularity of the GitHub projects? & \cite{fan2021makes} 
\\ 

\textbf{I2:} What are the code smells related to PTM in the downstream projects, and how do they affect theses projects?
& \cite{palomba2018beyond, zhang2022codeSmellsforMLApplications, van2021prevalence, cardozo2023prevalence}
\\

\textbf{I3:} When PTMs are used in GitHub repositories, what are engineers' testing practices for the PTMs they add as dependencies? Is there any correlation between the tests of the PTM by its maintainers, and the tests of the PTM by the downstream users? Do practices vary based on the purpose of the PTM, \eg computer vision vs. natural language processing? How do PTM downstream users deal with flakiness when testing a PTM?
& \cite{Li2022TestingMLSystemsinIndustry, Braiek2020onTestingMLPrograms,nahar2023dataset, eck2019understandingFlakyTest} \\

\textbf{I4:} Updating dependencies is a core software engineering activity. Suppose a GitHub repository depends on a PTM. How often does the GitHub repository update the dependency when that PTM is changed, \eg due to (1) PTM deprecation, (2) PTM improvement via a new version, or (3) PTM being made outdated by the release of a new model? What is the typical lag time for such updates? & \cite{hora2018developers, wan2021MLCloudAPIs} 
\\

\textbf{I5:} Software engineers often communicate through filing issue reports. What are the characteristics of issue reports on the PTM packages, \eg in terms of the kinds of questions asked, responsiveness of maintainers, issue density, and issue staleness? How often does the topic of reproducibility come up (cf. the ``ML reproducibility crisis'')? How do these attributes differ from the characteristics of issue reports in GitHub repositories? & 
\cite{jiang2023CVReengineering, yang2023AIRepoIssues} \\

\textbf{I6:} When engineers deploy software applications that make use of PTMs, they may prefer to use a deployment framework, \eg the ONNX RunTime, rather than a development framework such as PyTorch. Which of the several competing deployment frameworks (ONNX RunTime, MM-DNN, NNET, TFLite, etc.) is the most popular, and is there any evidence of why? Do GitHub users make the transformation to deployment themselves or do the PTM authors provide the deployment-ready version? & \cite{Liu2023MLlibRecommendationbasedonKnowledgeGraph, zhang2022teeslice, velasco2018optimumSelectionofDNNModelsandFrameworks}
\\

\bottomrule
\end{tabular}
\end{table*}
}

\clearpage

\appendices


\section{Dataset schema} \label{sec:Schema}
\begin{figure*}[b!]
    \centering
    \includegraphics[width=1\textwidth]{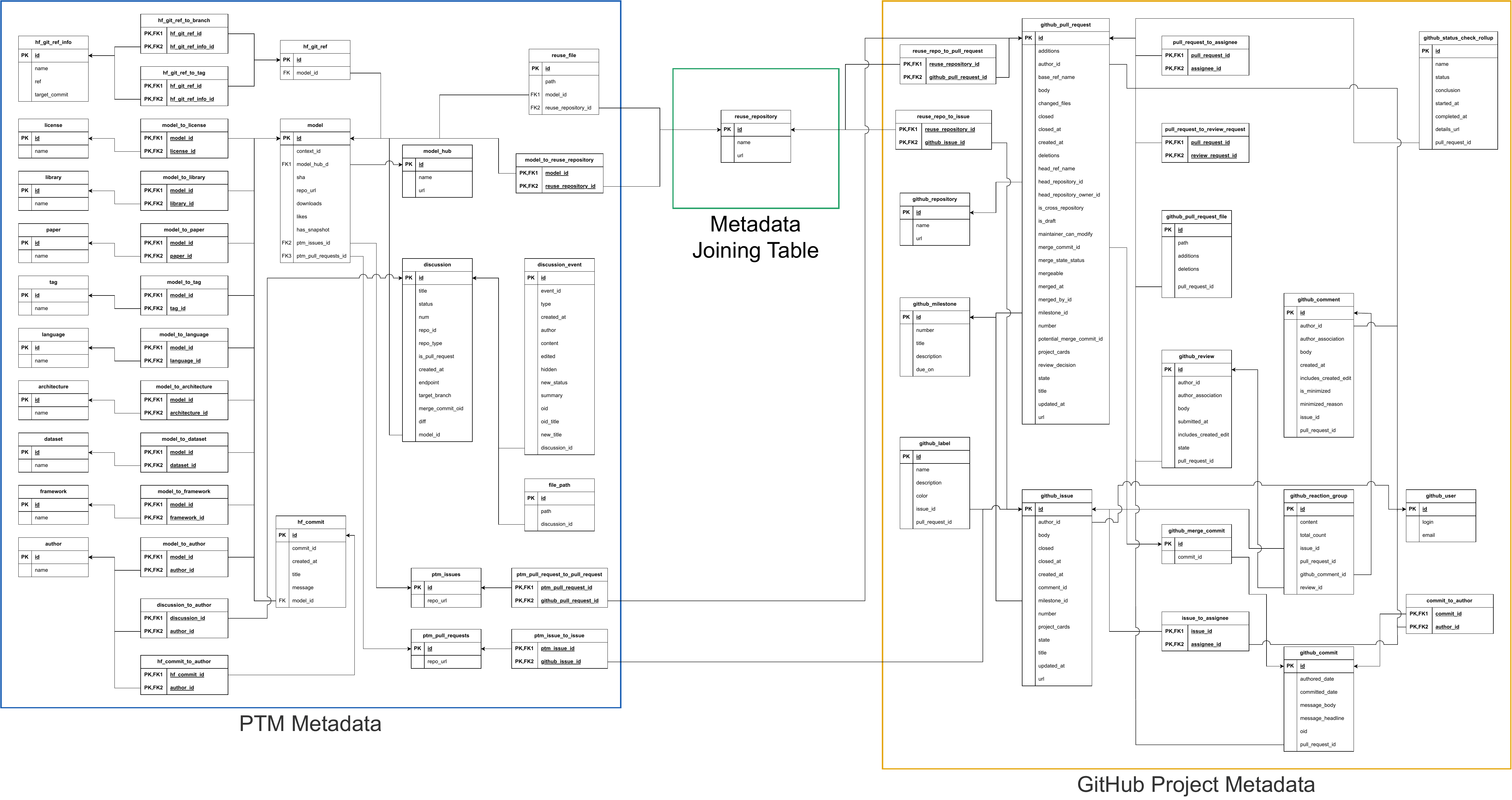}
    \vspace{0.3cm}
    \caption{
    \small
    The \DatasetNickname data schema. 
    The database has three regions: one set of tables for PTMs, one set of tables for GitHub projects, and one table linking the two.
    The underlying PTMs and GitHub repositories are stored in the Globus share and can be fetched on demand.
    A more navigable version of the schema is available in the demo repository.
    }
    \label{fig:DataSchema}
\end{figure*}

\balance
The detailed schema is shown in~\cref{fig:DataSchema}.
The definition in SQL is available to miners.

\clearpage

\section{Data collection}







\subsection{PTMs}

\subsubsection{What is a PTM and PTM package?}
\begin{figure}[h]
    \centering
    \includegraphics[width=0.9\linewidth]{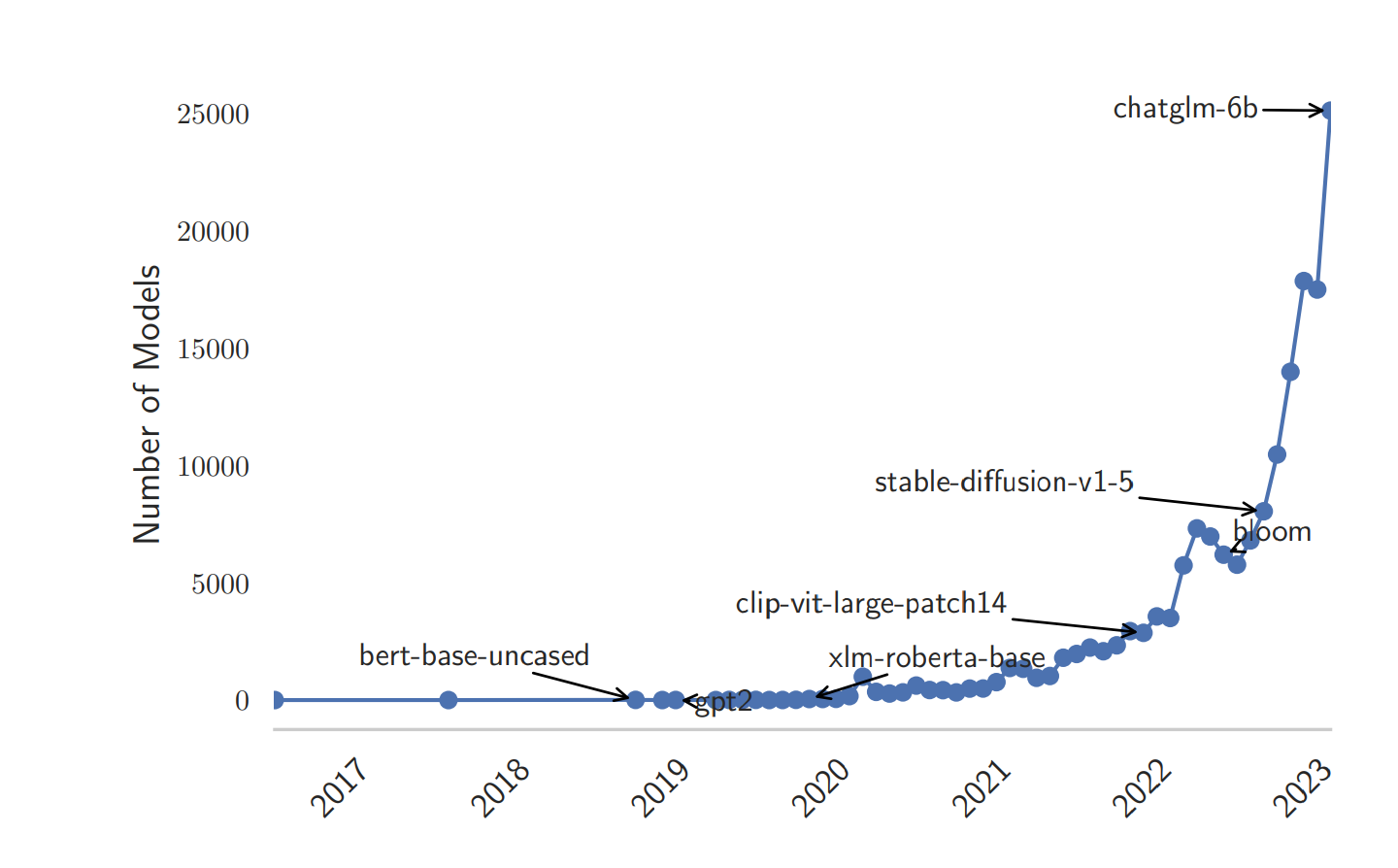}
    \caption{Evolution of the total number of Hugging Face models per month. This figure is reused from \cite{castano2023exploring}.}
    \label{fig:PTMPopularity}
\end{figure}

Pre-trained deep learning models (PTMs) are often shared through deep learning model registries, such as Hugging Face. Engineers can directly reuse these PTMs, or fine-tune them for specific downstream tasks. A PTM package typically includes a model card (a form of README), as well as the model's architecture and weights~\cite{Jiang2022PTMReuse}.
In recent years, the popularity of PTMs has been steadily rising~\cite{kathikar2023assessing, castano2023exploring}. 
As illustrated in \cref{fig:PTMPopularity}, the total number of Hugging Face models has seen a consistent increase on a monthly basis.
Recent work shows increasing interest from the software engineering mining community in PTMs~\cite{Jiang2022PTMSupplyChain, Jiang2022PTMReuse, davis2023PTMReuseChallengesandDirections, ait2023hfcommunity}. 
These works have identified the potential mining data that the community can take advantage of.
In the past year the first mining efforts of PTMs and software engineering practices have begun~\cite{castano2023exploring, kathikar2023assessing}.

\subsubsection{PTM Collection}
Our PTM data collection includes three parts:
    (1) We saved \NEWnumberOfPTMRepos PTM snapshots. This included the most popular PTM packages (\ie with over 10K downloads) on Hugging Face, which resemble a ``git-like'' structure, and all PTMs on PyTorch. This part of the data can provide a comprehensive view of PTM packages.
    (2) Among these ``full'' metadata, \TotalNumberOfLinks links from the PTMs to the downstream GitHub repositories have been identified. This part of the data can be connected to downstream GitHub data and allow miners to analyze the relationship between them.
    (3) For all PTMs hosted on Hugging Face and PyTorch, we retrieved their metadata, resulting in a total number of \TotalNumberOfPackagesMetadata PTM package metadata being included in \DatasetNickname. The miners can answer research questions based on metadata only, such as analyzing PTM naming conventions.


\subsubsection{Soundness and Completeness}
\DatasetNickname is comprehensive in terms of popular PTM packages, as it includes snapshots of those with over 10,000 downloads on Hugging Face. This provides a full view of widely-used PTMs and their connections to downstream GitHub projects, facilitating in-depth analysis. Additionally, the dataset includes metadata from all other PTMs on Hugging Face, which can be used for metadata-based analyses. 
\DatasetNickname enhances the diversity of PTM data by incorporating PTM packages from PyTorch Hub, including all available model repositories and their associated pull requests and issues.

\subsubsection{Implementation} \label{sec:PTM_Implementation}
\hspace{0mm}
Metadata is collected using an \textit{Extract-Translate-Load (ETL)} pipeline for each model hub.
The ETL pipeline can be generalized to the following steps:

\begin{itemize}
    \item \textbf{Extract}: Obtain metadata that is available from each model hub's API.
    \item \textbf{Transform}: Use metadata to collect more information about PTMs (\ie examine Github Metadata for a linked repository) as well as download PTM package of GitHub repository snapshot. Transform the data into intermediate representation to simplify Loading. 
    \item \textbf{Load}: Load the transformed data into a database for long-term storage and retrieval
\end{itemize}


Each model hub has a unique implementation of the Extract stage, but the functionality is the same.
\begin{itemize}
    \item \textbf{Hugging Face}: PTM package metadata is downloaded using the \code{HuggingFace\_hub} Python library.
    \item \textbf{PyTorch}: The markdown files in the root of the PyTorch GitHub repository, which correspond to each PTM package's repository, are downloaded and subsequently scraped for metadata.
\end{itemize}

During the Transform stage, data that matches the \DBFileName metadata schema is transformed into an intermediate representation, while data that doesn't match the schema is transformed into a JSONB blob.
This is to allow for both consistency across model hubs, as well as maintaining hub specific metadata.

\subsection{Mapping GitHub projects to PTMs} \label{sec:DataCollection-Linking}
\balance
To evaluate PTM usage within projects, a PTM must be mapped back to a GitHub project.
While it is possible to use Hugging Face and PyTorch Hub hosted projects with libraries outside of the Python ecosystem, we filtered on GitHub projects that utilize Python libraries to interface with PTMs as the vast majority of projects we found utilized Python libraries.   
As Hugging Face and PyTorch hub do not provide metadata or a method to retrieve PTM usage within projects, by analyzing the source code of a GitHub project, it is possible to map the two. The whole methodology has been described below:

\subsubsection{Signature Collection}
\begin{figure}[h]
    \centering
      \fbox  {
    \includegraphics[width=0.95\linewidth]{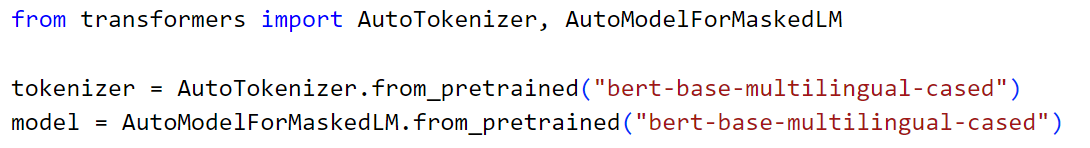}}
    \caption{An Example Code Snippet to Use PTMs from Hugging Face Hub}
    \label{fig:usage_HF}
\end{figure}

For leveraging Hugging Face PTMs, dedicated functions/methods specific to each library are available. These functions/methods allow loading PTMs by inputting their names as arguments.
In this step, we manually retrieve the libraries, along with their corresponding import identifiers and the necessary function/method names essential for PTM loading. 
This compilation process is guided by Hugging Face's official documentation.\footnote{\url{https://github.com/HuggingFace/hub-docs/blob/main/js/src/lib\\/interfaces/Libraries.ts}}
The culmination of import identifiers and the essential function/method names is referred to as a Signature. 
For example, when accessing the PTMs provided by the Diffusers library, the corresponding signature encompasses \code{diffusers} and \code{from\_pretrained}.
Note that some libraries offer multiple methods to load PTMs, and we have taken all of these methods into account. 
For instance, the \code{transformers} library presents two distinct approaches: \code{from\_pretrained} and \code{pipeline}, both of which can be utilized to load PTMs. Figure~\ref{fig:usage_HF} shows an example code snippet to load the PTMs provided by \code{Transformers} library.

Throughout this process, we exclude libraries not visible on the website (such as \code{k2}, \code{doctr}, \code{mindspore}), as well as those unsuitable for downstream projects. 
An example of such incompatibility is when PTMs can only be used via command line or download, or when they lack the "use in library" feature\footnote{\url{https://HuggingFace.co/facebook/musicgen-large}}. 
After filtering, a total of \GitHubReuseLibrariesHF libraries remain in the compilation.

\begin{figure}[h]
    \centering
      \fbox  {
    \includegraphics[width=0.95\linewidth]{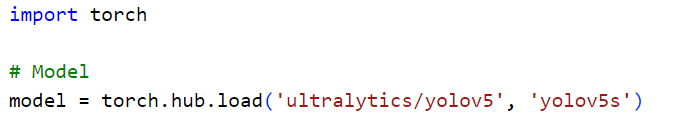}}
    \caption{An Example Code Snippet to Use PTMs from PyTorch Hub}
    \label{fig:usage_PT}
\end{figure}

To use PyTorch PTMs, there are two approaches: (1) \code{torch.hub.load} function that provides a way to load PTMs by specifying their names as arguments. and (2) Alternatively, one can utilize library-specific classes provided by \code{torchvision} or instances provided by \code{torchaudio} and \code{torchtext}. These classes and instances are tailored to represent individual PTMs. The first approach and alternative approach provided by the \code{torchvision} library contain keyword based parameters i.e., pretrained/weights to control the model usage with pretrained weights or random initialization.
The above mentioned two approaches give us a total of \GitHubReuseSignaturePT signatures. Figure~\ref{fig:usage_PT} shows an example code snippet to load the PTMs provided by the \code{PyTorch} library.

\vspace{-5mm}

\subsubsection{Preliminary repository collection based on Sourcegraph Search}
The subsequent task involves locating these signatures within the code. Although we attempted to use GitHub search, it does not facilitate a comprehensive search as it provides the top 1000 results.
Instead, we utilize the Sourcegraph command line interface, \code{src}\footnote{\url{https://docs.sourcegraph.com/cli}}, to detect projects containing pertinent files that employ the gathered signatures to interact with Hugging Face and PyTorch Hub. 

Our search pattern incorporates the signatures gathered earlier, and we match them against the content of files within GitHub repositories. Our search criteria encompass repositories that are not archived (default behavior), not forked (default behavior), and publicly accessible, specifically focusing on Python files. 
For example, a query for \code{Diffusers} is structured as \code{``src select:file visibility:public count:all lang:Python content:`from diffusers' AND from\_pretrained(''}. Our search query accommodates both `from import' and `import' statements. Our search results include the corresponding code snippets, file names, and repository names. The projects were identified during July 5-12. 
Based on the count of the repositories, we select the top 5 Hugging Face libraries for our data collection including \code{Transformers}, \code{Spacy}, \code{Sentence-Transformers}, \code{Diffusers}, and \code{Timm}. For PyTorch, we consider all of the corresponding signatures. Our dataset comprises well-recognized GitHub repositories with an average star count of \GitHubReuseRepoSourceGraphAvgStar.


We have obtained local copies of the GitHub repositories by using the git clone command. This process took approximately 12 days to complete and resulted in the download of \GitHubReuseRepoCountSourceGraph repositories, collectively amounting to a total size of \GitHubReuseRepoCountSourceGraphSize. 

\subsubsection{Extracting PTMs via Static Analysis}
As Sourcegraph's search feature relies on text-based patterns, the possibility of encountering false positive results exists. 
To mitigate this concern, we perform static analysis on GitHub repositories with the Scalpel framework~\cite{li2022scalpel}. 
For each relevant source code file associated with a specific function signature, we construct an abstract syntax tree and extract the function calls contained within the file. 
Subsequently, we retrieve the complete and qualified names of each identified function call and cross-reference them with our predefined signatures which gives us a total of \GitHubReuseRepoCountTP repositories. 
Additionally, we go a step further by extracting both the positional and keyword arguments that are associated with the function calls that match our target signatures. 
Our analysis is equipped to capture any argument that possesses a static value. 
We then utilize the list of PTMs from PTM Torrent V2 to identify the repositories that statically call PTMs which gives us a total of
\GitHubReuseRepoCountStaticMatch repositories. We store the corresponding repositories and files for each of the matched PTMs. It is important to note that a single repository can utilize multiple PTMs, and similarly, a single PTM can be employed across multiple repositories. 

\subsubsection{Soundness and Completeness of Collected Repositories}
\begin{figure}[h]
    \centering
    \includegraphics[width=0.9\linewidth]{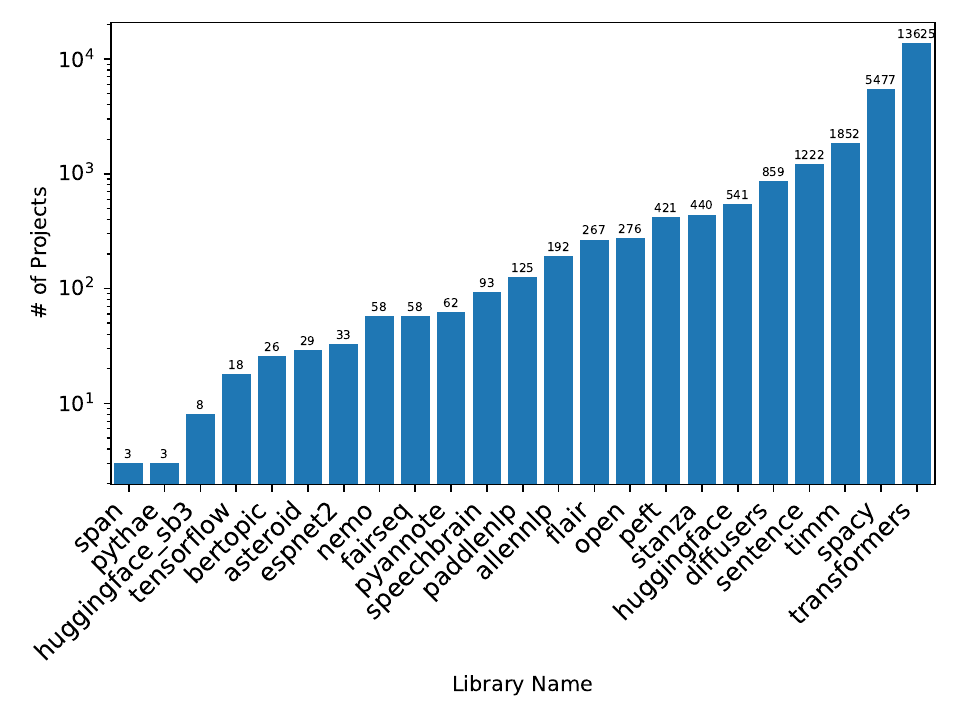}
    \caption{Number of projects that use a specific library as captured via Sourcegraph code search}
    \label{fig:NumberOfProjectsperLibrary}
\end{figure}

For the PTMs hosted on the Hugging Face hub, our dataset provides usage considering the five libraries, i.e., \code{Transformers}, \code{Spacy}, \code{Timm}, \code{Sentence-Transformers}, and \code{Diffusers}.
These libraries were chosen because they comprise the top five libraries used in GitHub projects as shown in \cref{fig:NumberOfProjectsperLibrary}. For the PTMs from the PyTorch hub, we did not filter by the library. Our dataset comprises \code{torchvision}, \code{torchaudio}, \code{torchtext}, along with PyTorch hub.

Static analysis was carried out due to the limitations of the text search conducted using Sourcegraph. We resolve the fully qualified names for each function call to accurately identify True Positive results. This results in a total of \GitHubReuseRepoCountTP repositories that genuinely contain the practical utilization of the PTMs. Our dataset encompasses projects created up until \GitHubReuseRepoSourceGraphDate.

\subsubsection{Extracting GitHub Issues and Pull Requests}
By analyzing the discussions that community members have about the project within the project's issue tracker, it is possible to identify not only PTMs of interest w.r.t the project, but the potential future direction of a project w.r.t the techniques implemented by the PTM.

To collect the issues and pull requests associated with the GitHub repositories, we use GitHub's official command line interface \code{gh}.
We consider all states (\ie open and closed) while collecting the issues and pull requests associated with each repository. 
Each of the issue and pull request metadata responses contain all available relevant fields provided by GitHub CLI. 
Specific response fields are listed in ~\cref{tab:GHCLI_RequestFields}. 
Example commands to retrieve relevant data can be found in \DatasetNickname GitHub repository.
%
%
%
Targeting \GitHubReuseRepoCountTP repositories, we retrieve the issues or pull requests resulting in  \GitHubReuseIssueCount repositories with issues and \GitHubReusePrCount repositories with pull requests.

Altogether, our dataset encompasses a total of \GitHubReuseRepoCountInDB repositories, which involve occurrences of issues, pull requests or static utilization of PTMs.

{
\renewcommand{\arraystretch}{0.5}
\begin{table}[h]
\centering
\caption{
    JSON response fields captured when collecting issues and pull requests for \GitHubReuseRepoCountTP GitHub repositories.
    }
\begin{tabular}{p{0.33\linewidth}p{0.57\linewidth}}
\hline
\toprule
    \textbf{Request Type} & \textbf{Response Fields} \\                                                                              

\midrule
    Issue Metadata & assignees, author, body, closed, closedAt, comments, createdAt, id, labels, milestone, number, projectCards, reactionGroups, state, title, updatedAt, url \\

\midrule 
    Pull Request Metadata & additions, assignees, author, baseRefName, body, changedFiles, closed, closedAt, comments, commits, createdAt, deletions, files, headRefName, headRepository, headRepositoryOwner, id, isCrossRepository, isDraft, labels, maintainerCanModify, mergeCommit, mergeStateStatus, mergeable, mergedAt, mergedBy, milestone, number, potentialMergeCommit, projectCards, reactionGroups, reviewDecision, reviewRequests, reviews, state, statusCheckRollup, title, updatedAt, url \\

\bottomrule
\label{tab:GHCLI_RequestFields}
\end{tabular}
\end{table}
}

\section{Data Access Examples} \label{sec:Access_Examples}
To answer several of the proposed research questions in \cref{table:RQs}, we have released examples on how to interface with \DatasetNickname.
ORM methods and SQL examples for interfacing with the \DBFileName database are provided.
Code snippets for these examples are made available via the \DBAccessExamplesPath filepath in our GitHub repository.

\raggedbottom
\vspace{1cm}
\balance

\bibliographystyle{bibliography/IEEEtran}
\balance
\bibliography{bibliography/DualityLab, bibliography/WenxinZotero, bibliography/Reference}

\end{document}